\documentclass[preprint2]{aastex}
\shortauthors{Nu\~nez et al.}

\begin{document}


%\title{Symbiotic stars in X-rays III: long term variability.}
\title{Symbiotic stars in X-rays III: {\em Suzaku} observations.}
% * <jeno@astro.columbia.edu> 2015-08-14T23:03:02.901Z:
%
%  I like this title better than the one in the response to the referee.  :)
%
% ^ <jeno@astro.columbia.edu> 2015-08-14T23:03:38.936Z.


%\author{Natalia, Tommy, Koji, Jeno, Juan}

\author{N. E. Nu\~nez\altaffilmark{1}, T. Nelson\altaffilmark{2}, K. Mukai\altaffilmark{3}, J. L. Sokoloski\altaffilmark{4} and G. J. M. Luna\altaffilmark{5} }

\altaffiltext{1}{Instituto de Ciencias Astron\'omicas de la Tierra y del Espacio (ICATE-UNSJ, CONICET), Av. Espa\~na (S) 1512, J5402DSP, San Juan, Argentina}
\email{nnunez@icate-conicet.gov.ar}

\altaffiltext{2}{Minnesota Institute for Astrophysics, University of Minnesota, Minneapolis, MN, 55455, USA}
\altaffiltext{3}{CRESST and X-ray Astrophysics Laboratory, (NASA/GSFC), Greenbelt MD 20 771, USA. Department of Physics, University of Maryland, Baltimore County, 1000 Hilltop Circle, Baltimore, MD, 21 250, USA)}
\altaffiltext{4}{Columbia Astrophysics Lab, 550 W120th St., 1027 Pupin Hall, MC 5247 Columbia University, 10027, New York, USA}
\altaffiltext{5}{Instituto de Astronom\'ia y F\'isica del Espacio (IAFE, CONICET-UBA),
	Av. Inte. G\"uiraldes 2620, C1428ZAA, Buenos Aires, Argentina}

%*****************************


\begin{abstract}

% * <jeno@astro.columbia.edu> 2015-04-20T20:08:32.933Z:
%
%  I think it would be great if we could add a sentence motivating this research to the beginning of the abstract.
%
We describe the X-ray emission as observed with {\em Suzaku} from five symbiotic stars that we selected for deep {\em Suzaku} observations  after their initial detection with {\em ROSAT, ASCA} and {\em Swift}. We find that the X-ray spectra of all five sources can be adequately fit with absorbed, optically thin thermal plasma models, with either single- or multi-temperature plasmas. These models are compatible with the X-ray emission originating in the boundary layer between an accretion disk and a white dwarf.  The high plasma temperatures of  kT~$>3$~keV for all five targets were greater than expected for colliding winds.
Based on these high temperatures, as well as  previous measurements of UV variability and UV luminosity, and the large amplitude of X-ray flickering in 4~Dra, we conclude that all five sources are accretion-powered through predominantly optically thick boundary layers. Our X-ray data allow us to observe a small, optically thin portion of the emission from these boundary layers. Given the time between previous observations and these observations, we find that the intrinsic X-ray flux and the intervening absorbing column can vary by factors of three or more on a time scale of years. However, the location of the absorber and the relationship between changes in accretion rate and absorption are still elusive.

\end{abstract}



\section{Introduction}

When observed at optical wavelengths, symbiotic stars (SS) show a composite spectrum that suggests that they are binary systems. A hot, compact component (usually a white dwarf, WD) contributes to the blue-UV region of the spectrum, while a cool red giant dominates the spectrum at longer wavelengths. 

Observations at other wavelengths reveal a very complex and rich scenario for these systems. Optical, infrared and UV spectral regions are rich in emission lines from forbidden and permitted transitions, which arise mainly from photoionization and recombination of the nebular plasma heated by the the hot component \citep{kenyon09}. Radio, optical, and X-ray observations reveal jets with velocities of a few hundred to around 1,000~km~s$^{-1}$ \citep[e.g.,][]{brocksopp04,crocker01,kellogg07} and thermal emission from the ionized red-giant wind \citep{seaquist90,seaquist93}. Symbiotic stars can even produce $\gamma$-rays during nova eruptions \citep[e.g. V407~Cyg,][]{abdo2010,v407cyg}. Symbiotics are now recognized as a population of X-ray sources. From the $\sim$~220 systems known \citep{bel00}, 45 have been detected at X-ray wavelengths, most of them with emission in the 0.3 to 10 keV range. A few, however, were detected at energies up to 100 keV \citep{kennea09}. 

The X-ray emission from symbiotic stars  consist of some combination of four distinct spectral components (dubbed $\alpha, \beta, \gamma$ and $\delta$), which most likely arise from distinct emission regions and/or processes \citep[][hereafter Paper I]{Paper1}. In particular, $\alpha$, $\beta$ and $\delta$-type X-ray spectral components come
from thermal X-ray emission (optically thin or blackbody-type) that arises from the quasi-stable nuclear burning on the WD surface ($\alpha$), a colliding-wind region with kT~$\lesssim$ 1 keV ($\beta$), or an accretion-disk boundary layer ($\delta$). Those symbiotics with a neutron star as the accreting compact object and non-thermal, power-law type X-ray spectra are classified as $\gamma$-type by \cite{murset97}.

Sensitive and broadband X-ray satellites such as {\em Suzaku} have played a significant role in observing symbiotics, specially those with X-ray emission above 10 keV \citep[T CrB, CH Cyg, V648 Car;][]{luna08, mukai07, kennea09}. \citet{Paper1} and \citet{nunez14} studied and classified the first X-ray detections of many symbiotics with Swift, {\em XMM-Newton} and/or {\em Chandra}. Here we describe {\em Suzaku} observations of CD-28~3719, Hen~3-1591, Hen~3-461, EG~And and 4~Dra. All five symbiotic stars have previously been observed with {\em Swift}, {\em ROSAT} or {\em ASCA} \citep{Paper1,murset97}; the {\em Suzaku} observations provide higher quality X-ray spectra. Here we present the analysis of these {\em Suzaku XIS} data. In Section \ref{sec:obs-red} we detail the data reduction and analysis for these five sources, while in Section \ref{sec:results} we present the results. We discuss our interpretation of the results in Section \ref{sec:disc} and explain our conclusions in Section \ref{sec:concl}.

\section{Observations and data reduction}
\label{sec:obs-red}

{\em Suzaku} observed the five symbiotic stars with the X-ray Imaging Spectrometer \citep[XIS;][]{koyama07}. Details of each observation are presented in Table \ref{table:info1}. Data were taken with the XIS0, XIS1, and XIS3 detectors, which are sensitive in the 0.2-12 keV range (XIS2 has not been operational since November 2006). All sources were too faint to be detected with the Hard X-ray detector (HXD).  We reprocessed all data using the \texttt{aepipeline} script and obtained event files with the processing version 2.5.16.29 (2014-07-01) applied.  

Source spectra and light curves were extracted from circular regions centered on the source SIMBAD\footnote{http://simbad.u-strasbg.fr/simbad/sim-fid} coordinates. The recommended radius of the extraction region\footnote{http://heasarc.gsfc.nasa.gov/docs/suzaku/analysis/abc/} is 260$\arcsec$ (this size encircles 99\% of the point source flux); given the source and background brightnesses, however, we were able to use this size only in the case of 4~Dra. Comparing the source spectra with that of the background, we found that the optimal radius for the source region, which maximizes the signal-to-noise ratio, was of 120$\arcsec$ in the case of Hen~3-1591, CD~-28~3719, and EG~And, and 60$\arcsec$ for Hen~3-461. Background spectra and light curves were extracted from annular regions centered on the source (with inner and outer radius of 340$\arcsec$ and 430$\arcsec$, respectively) in the case of Hen~3-1591, CD~-28~3719, Hen~3-461, and 4~Dra, while a circular region with 160$\arcsec$ radius was used for EG~And 
because the location of the source on the chip did not allow us to
select an annular region for the background. The response matrices
were created using the \texttt{xisrmfgen} and \texttt{xisarfgen}
tools. We then fit the binned spectra (grouped by a minimum of 20 to
25 counts per bin) using
XSPEC\footnote{http://heasarc.gsfc.nasa.gov/docs/xanadu/xspec/} and
the $\chi^{2}$ statistic to select the best-fit models.

%We found that 
One of our targets, Hen~3-1591, was observed serendipitously with {\em
  ASCA} \citep{tanaka94} on 1999 September 22 during an observation of
the supernova remnant (SNR), \object{G5.2-2.6}. The Solid-state
Imaging Spectrometer (SIS) was operated in 1-CCD mode for this
observation (of the 4 chips available), which put Hen~3-1591 outside
the operational SIS field of view. On the other hand, Hen~3-1591 was
securely in the field of view of the Gas Imaging Spectrometer (GIS)
instrument, which has two units (GIS2 and GIS3). We selected intervals
when the satellite was outside the South Atlantic Anomaly (SAA), when
the attitude control was stable and the satellite pointed within
0.02 degrees of the target, and when the line of sight was greater
than 5 degrees above the Earth limb. We also applied standard
selection 
%expression 
criteria combining monitor count rates and geomagnetic cutoff
rigidity (COR) to exclude time intervals of high-particle background,
in the end obtaining 15 ks of good on-source data.

For the {\em ASCA} observation on Hen~3-1591, we extracted the source
spectrum from a 6$^{\prime}$-radius circular extraction region
centered on the source, and the background spectrum from a 6$^{\prime}$-radius region centered at $\alpha$=271.6619, $\delta$=-25.8192, away from the SNR and other obvious sources. We used the v4.0 rmf downloaded from the CALDB and generated an arf file for each unit of the GIS. We then combined the GIS2 and GIS3 spectra and responses, ignored the data outside the well-calibrated range of 0.7--10 keV, and binned the data by a factor of 32, leaving 26 channels.

  \begin{deluxetable}{l r c c}
%\rotate
   \tabletypesize{\footnotesize}
  \tablecolumns{4}
 \tablewidth{0pt}
  \tablecaption{Observation Log \label{table:info1}}
  \tablehead{\colhead{Source} & \colhead{Date} & \colhead{ObsId}  & \colhead{Exp. Time [ks]}\\}   % \colhead{Coordinates (J2000)} & \colhead{Search-Offset [arcmin]}} 
  \startdata
  \hbox to 0.9in{CD~-28~3719\leaders\hbox to 0.8em{\hss.\hss}\hfill}   & 2013-10-12 & 408032010 & 14 \\ % 07 01 08.78 -29 07 00.5  &  & 0.597 \\
  \hbox to 0.9in{Hen~3-1591\leaders\hbox to 0.8em{\hss.\hss}\hfill}     & 2012-10-03 & 407042010 & 51  \\ % 18 07 32.71 -25 53 53.2  & & 0.226\\
  \hbox to 0.9in{- {\em ASCA} \leaders\hbox to 0.8em{\hss.\hss}\hfill} & 1999-09-22 & 57055000  & 15 \\
  \hbox to 0.9in{Hen~3-461\leaders\hbox to 0.8em{\hss.\hss}\hfill}      & 2012-12-17 & 407007010 & 46 \\ % 10 39 06.70 -51 24 31.7  &  & 0.447  \\ 
  \hbox to 0.9in{EG~And\leaders\hbox to 0.8em{\hss.\hss}\hfill}        & 2011-02-05 & 405034010 & 100  \\ % 00 44 40.32 +40 40 22.8 & & 0.708\\
  \hbox to 0.9in{4~Dra\leaders\hbox to 0.8em{\hss.\hss}\hfill} & 2010-04-18 & 405035010 & 42 \\ %& 12 30 06.66 +69 12 04.06 &  & 1.029 \\
                                                                          & 2011-11-09 & 406041010 & 42 \\ % 12 30 06.66 +69 12 04.06 &  & 1.201 \\
 \enddata
  \end{deluxetable}

In those cases where the thermal nature of the emission was not obvious, i.e. lines arising from optically thin thermal emission were weak or absent, we evaluated whether any of three spectral models --- an absorbed single-temperature optically thin thermal plasma, an absorbed multi-temperature cooling flow, or an absorbed non-thermal plasma --- properly fit the data based on their $\chi^{2}_{\nu}$. The parameters of the best-fit models are listed in Table \ref{tab:models}. We used the \texttt{Tbabs} to model absorption that completely covered the sources of X-ray emission, using the abundances of \citet{wilms} and the cross-sections of \citet{verner}.

Once the best model was found, we used the unbinned data and C-statistic \cite{cash} to calculate the uncertainties in the parameters of the models and the flux. All errors
in the fit parameters correspond to 90\% confidence intervals (see Table \ref{tab:models}).

\section{Results}
\label{sec:results}

\subsection{Spectral analysis}

\begin{figure*}
 \includegraphics[]{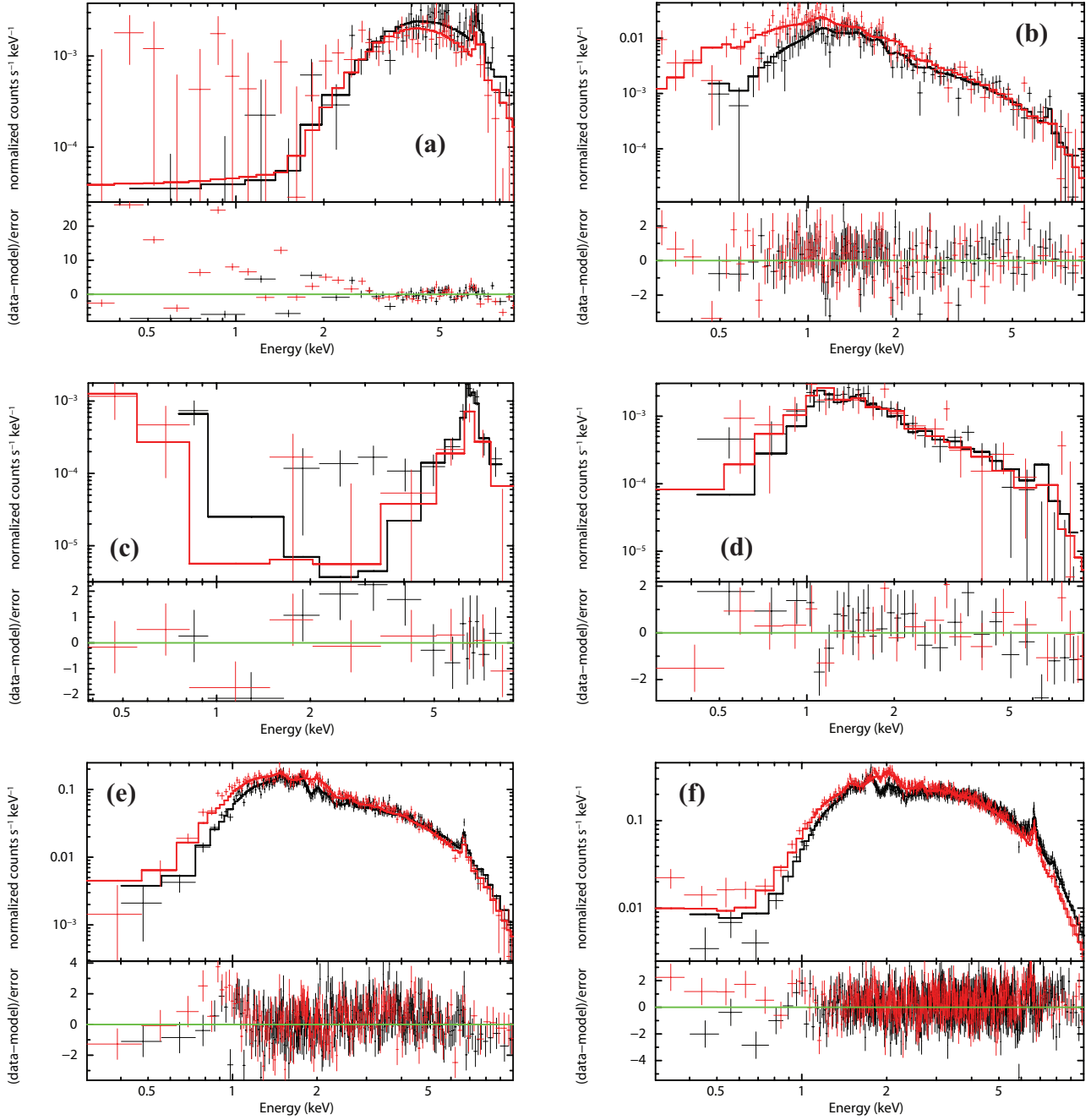}
 \caption{{\em Suzaku/XIS} spectra of ($a$) CD~-28~3719, ($b$)
   Hen~3-1591, ($c$) Hen 3-461, ($d$) EG~And and ($e$) 4~Dra ObsID
   4050 and ($f$) 4~Dra ObsID 4060. The solid lines show the best-fit models described in Section \ref{sec:results}. Red (XIS1), black (XIS0+3).} 
\label{fig1}
\end{figure*}

The X-ray emission from each of the five sources we observed with {\em
  Suzaku} was successfully modeled as absorbed, optically thin thermal
emission from either a single or multi-temperature
plasma. Temperatures were high, with {\em kT} of about a few keV for
all sources.
%, and the hardness of the spectra less than in previous observations,
%indicating lower absorbing columns. 
The fluxes from both objects in common with Paper 1 (Hen~3-461 and
CD~-28~3719) were higher when observed with {\em Swift} than when
observed more recently with {\em Suzaku}. The parameters from the best fit models are listed in Table \ref{tab:models} while spectra are shown in Fig. \ref{fig1}.

\subsubsection{CD~-28~3719}
\label{sec:rescd-28}

To improve the basic spectral modeling, we intended for our {\em  Suzaku} observation to provide a spectrum with a higher
signal-to-noise ratio than previous datasets. During the observation of CD~-28~3719, however, there were some problems in the acquisition of the XIS0 chip data, which did not return to the 5$\times$5 editing mode after dark frame dump during a South Atlantic Anomaly passage, and the data during these segments were corrupted. For this reason, we only analyzed the data from the XIS1 and XIS3 chips.  

The presence of excess counts in the $\sim$6.6 keV region led us to test optically thin thermal plasma models for the spectrum instead of non-thermal models, which should not produce emission lines in this spectral region. Fe K$\alpha$ fluorescent emission line, if present, should be centered at $\approx$ 6.4 keV. We interpret the line near
  6.6~keV as being due to a combination of H-like, He-like, and fluorescent Fe lines.
The best fit model for this source, whose x-ray spectrum we show in Figure~\ref{fig1},  consists of an absorbed,
single-temperature, optically thin thermal plasma with variable abundance, \texttt{Tbabs$\times$apec}. The  metal abundance was
0.44$^{+0.30}_{-0.17}$. The other best-fit model parameters are listed in Table \ref{tab:models}. The results from this fit are
commensurate with the results obtained from the {\em Swift} data analyzed in Paper I. 

\subsubsection{Hen~3-1591}
\label{sec:reshen315}

\begin{figure}
 \includegraphics[scale=0.25,angle=-90]{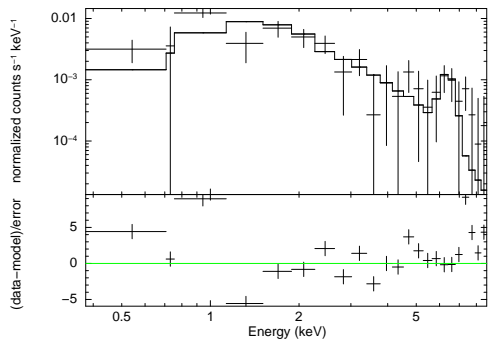}
 \caption{{\em ASCA}/GIS spectra of Hen~3-1591. The solid line shows the best-fit model described in Section \ref{sec:results}. } 
\label{fig:asca}
\end{figure}

Hen~3-1591 was serendipitously observed with {\em ASCA/GIS}, and
  our reduction of the {\em ASCA} data revealed that it was detected
  with a net count rate of 0.016 c/s/GIS.  
%The {\em ASCA} spectrum, however, was not of high enough statistical
%quality (see Fig. \ref{fig:asca}) to draw definitive conclusions (only
%$\sim$200 photons were detected), although some inferences can be
%drawn.  
Figure~\ref{fig:asca} shows the {\em ASCA} spectrum of Hen~3-1591,
which contains $\sim$200 photons.
A power-law fit 
%prefers 
with a photon index near 2.4, with excess counts around 6.6 keV, provides an acceptable fit. Adding a Gaussian,
the line centroid was found to be near 6.6 keV with an equivalent width well in excess of 1 keV (calculated by including \texttt{eqwidth} in the
gaussian component of the model). 
%  we calculated this value applied the parameter eweer in the gaussian component.
As this line is probably due to a combination of H-like, He-like, and fluorescent Fe lines,
the X-ray emission should be modeled as optically thin thermal
emission with reflection adding the 6.6 keV line rather than a
  power-law. In fact, a cooling flow plus Gaussian model gives a
good description of the observed spectrum. The maximum temperature is
around 14 keV, iron is strongly overabundant ($\sim$ twice the  solar
value), and the equivalent width of the fluorescent line ($\sim$ 600
eV) also requires an overabundance of iron. 
%Taken at face value, 
This X-ray spectrum suggests that Hen~3-1591 hosts a WD
accreting matter that is overabundant in Fe. 

%Our {\em Suzaku} spectrum aimed to identify, in a higher
%signal-to-noise spectrum, the best spectral model. 
To model the {\em Suzaku} spectrum, we tested a non-thermal model like
that suggested by the early {\em ROSAT} data \cite{murset97} and
thermal models inspired by the {\em ASCA} data, which
suggested that the X-ray
emission is due to an optically thin thermal plasma.
The two thermal models we tested (a single temperature apec model and
 multi-temperature cooling flow model) returned similar
statistics ($\chi^{2}_{\nu}$=1.2).
%and similar model parameters. 
We prefer the \texttt{Tbabs$\times$apec} model.
As with the {\em ASCA data}, the {\em Suzaku} data show indicate an
overabundance of iron. 
Based on the fit statistic alone, we cannot rule out an absorbed non-thermal model ($\chi^{2}_{\nu}$=1.2). 
Because of the low number of photons detected in the Fe K$\alpha$ line
region of the Suzaku spectrum, the continued presence of the H-like,
He-like emission lines, which arise from thermal emission, 
%firmly established. 
was not immediately obvious.  We addressed the significance of a detection of emission lines in the Fe K$\alpha$ region by adding a gaussian line profile (centered at 6.66$\pm$0.23 keV) to the power-law model, which represents the underlying continuum, and comparing this model with a simple absorbed power-law model using the Likelihood Ratio Test (LRT, \citet{protassov} with a 1,000 simulations. In 88.3 \% of the simulations, a model with an emission line at 6.66 keV yields $\chi^{2}_{\nu}$ values smaller that the model without the emission line. We take this as evidence of the thermal nature of the emitting plasma. 
% , however  cannot be completely ruled out based on the fit statistic, $\chi^{2}_{\nu}$=1.11, 
We discuss additional evidence that supports a thermal origin for the
X-ray emission in Section~\ref{sec:disc}.

\subsubsection{Hen~3-461}
\label{sec:reshen3461}

Almost two and half years after the {\em Swift} observation, {\em
  Suzaku} observed Hen~3-461 and only detected 285 source photons,
i.e. an 8$\sigma$ detection after background subtraction. Given the
low number of photons detected, we were only able to perform crude spectral modeling ($\chi^{2}_{\nu}$=0.93; see Table \ref{tab:models}). We noticed that a weak soft component was detected at a low significance level (2$\sigma$). We thus decided to include a second component in our spectral model, being aware of the similarity with the well-known, two-spectral component $\beta/\delta$-type emission observed in few WD symbiotics (e.g.~NQ~Gem, V347~Nor, see Paper I). 
We observed a strong excess of counts in the Fe K$\alpha$ region,
  which is naturally explained if the emission originates in an optically thin thermal plasma. Thus, our best spectral model consisted of two optically thin thermal components and a Gaussian emission line at 6.4 keV to account for the fluorescent Fe K$\alpha$ line. The hard component was modified by full- and partial-covering absorbers (\texttt{apec+Tbabs*pcfabs$\times$(apec+gauss)}).

  \renewcommand{\arraystretch}{1.2}
\begin{deluxetable}{l l l l c c c c}
 \rotate
% \begin{sidewaytable}{h}
 \tabletypesize{\scriptsize}
% \tabletypesize{\footnotesize}
\tablecolumns{8}
\tablewidth{0pt}
\tablecaption{Spectral Models \label{tab:models}} 
\tablehead{\colhead{Object} & \colhead{Model} & \colhead{Count rate} & \colhead{N$_{H,22}$\tablenotemark{a}} & \colhead{kT\tablenotemark{b}}  & \colhead{F$_{X}$\tablenotemark{c}} & \colhead{L$_{X}$\tablenotemark{d}} & \colhead{$\chi^{2}_{\nu}$}/d.o.f. \\
 & & [10$^{-2}$ c s$^{-1}$] &  &  [keV] &  & &\\ 
 }
\startdata
\hline
\hbox to 0.6in{\object{CD~-28~3719}   \leaders\hbox to 0.2em{\hss.\hss}\hfill} &  \texttt{Tbabs$\times$apec} & 0.9$\pm$0.1  & 10$_{-1}^{+2}$ & 8$_{-2}^{+2}$ & 14$\pm$1 & 18$\pm$1 (d/1 kpc)$^2$ & 1.2/181 \\
                      &  \texttt{Tbabs$\times$mkcflow} & $\cdots$ & 13$_{-2}^{+3}$ & 11$_{-4}^{+9}$ & 19$\pm$1 & 22$\pm$1 (d/1 kpc)$^2$ & 1.2/182\\

%***************************************************************************************************************
\hline
\hbox to 0.6in{\object{Hen 3-1591} \leaders\hbox to 0.2em{\hss.\hss}\hfill} &  \texttt{Tbabs$\times$apec} & 3.7$\pm$0.1  & 0.04$\pm$0.01  & 3$\pm$1 &  6.0$\pm$0.2 & 6.1$\pm$0.2 (d/1 kpc)$^2$ & 1.2/210\\
&  \texttt{Tbabs$\times$mkcflow} & $\cdots$ & 0.10$\pm$0.04 & 6$\pm$1  & 6.9$\pm$0.3 & 7.8$\pm$0.3 (d/1 kpc)$^2$ & 1.2/214 \\%Mdot=6.98E-11

%***************************************************************************************************************
\hline
\hbox to 0.6in{\object{Hen 3-461}   \leaders\hbox to 0.2em{\hss.\hss}\hfill} &  \texttt{apec+Tbabs$\times$pcfabs$\times$(apec+gauss)} & 0.2$\pm$0.1  & 3$_{-2}^{+5}$  & 8$^{+5}_{-3}$  &  26$\pm$5 & 31$\pm$6 (d/1 kpc)$^2$ & 1.5/23 \\
&   & & 75$\pm$25; CF=0.97$\pm$0.02 & &   & & \\
%***************************************************************************************************************
\hline
\hbox to 0.6in{\object{EG And}   \leaders\hbox to 0.2em{\hss.\hss}\hfill} &  \texttt{Tbabs$\times$apec} & 0.4$\pm$0.1 & 0.3$_{-0.2}^{+0.9}$ & 3$\pm$2 & 1.1$\pm$0.2 & 0.4$\pm$0.1 (d/512 pc)$^2$ & 0.9/134 \\
         &  \texttt{Tbabs$\times$mkcflow} & $\cdots$ & 0.9$\pm$0.3 & 3$\pm$1 & 2.4$\pm$0.2 & 0.8$\pm$ 0.1 (d/512 pc)$^2$ & 1.1/134\\

%*************************************************************************************************************** 
\hline
1st scenario &&&&&& \\
(see Sect. \ref{sec:res4dra}) &&&&&& \\
\hline
\hbox to 0.6in{\object{{4~Dra}(4050)}   \leaders\hbox to 0.2em{\hss.\hss}\hfill}   &  \texttt{Tbabs$\times$pcfabs$\times$vapec} & 38$\pm$1 & 0.43$\pm$0.05 & 6.1$\pm$0.2 & 118$\pm$2 & 5$\pm$1 (d/190 pc)$^2$ & 1.1/5406 \\
 &  &  & 1.3$\pm$0.5 , CF=0.22$\pm$0.07 & &  &  &  \\

\hbox to 0.6in{\object{{4~Dra}(4060)}   \leaders\hbox to 0.2em{\hss.\hss}\hfill}   &  \texttt{Tbabs$\times$pcfabs$\times$vapec} &110$\pm$1 & 0.96$\pm$0.06 & 6.1$\pm$0.2 & 577$\pm$3 & 25$\pm$2 (d/190 pc)$^2$ & 1.1/5406\\
 &  &  & 2.7$\pm$0.2, CF=0.62$\pm$0.04 & &  &  &  \\

\hbox to 0.6in{\object{{4~Dra}(4050)}   \leaders\hbox to 0.2em{\hss.\hss}\hfill}&  \texttt{Tbabs$\times$pcfabs$\times$mkcflow} & $\cdots$ & 0.47$\pm$0.15 & 14$\pm$1 & 130$\pm$2 & 6$\pm$1 (d/190 pc)$^2$ & 1.1/5406 \\
 &  &  & 0.7$_{-0.2}^{+0.4}$, CF=0.45$_{-0.15}^{+0.09}$ & &  &  &  \\

\hbox to 0.6in{\object{{4~Dra}(4060)}   \leaders\hbox to 0.2em{\hss.\hss}\hfill}&  \texttt{Tbabs$\times$pcfabs$\times$mkcflow} & $\cdots$ & 1.32$\pm$0.04 & 14$\pm$1 & 640$\pm$3 & 27$\pm$1 (d/190 pc)$^2$ & 1.1/5406 \\
 &  &  & 2.6$_{-0.3}^{+0.4}$, CF=0.51$\pm$0.02 & &  &  &  \\
%\\
\hline
2nd scenario &&&&&& \\
% % % %  Esta parte ya esta, no tocar!!!
%Solo agreg?? los LX
(see Sect. \ref{sec:res4dra}) &&&&&& \\
\hline
\hbox to 0.6in{\object{{4~Dra}(4050)}   \leaders\hbox to 0.2em{\hss.\hss}\hfill} &  \texttt{Tbabs$\times$pcfabs$\times$vapec} & $\cdots$ & 0.47$_{-0.04}^{+0.03}$ & 4.9$\pm$0.3  & 124$\pm$1 & 5$\pm$1 (d/190 pc)$^2$ & 1.01/1794 \\
 &  &  & 2.6$_{-0.8}^{+1.1}$, CF=0.29$\pm$0.05 & &  &  &  \\

\hbox to 0.6in{\object{{4~Dra}(4060)}   \leaders\hbox to 0.2em{\hss.\hss}\hfill}&  \texttt{Tbabs$\times$pcfabs$\times$vapec} & $\cdots$ & 0.91$\pm$0.07 & 6.4$\pm$0.2 & 547$\pm$2 & 24$\pm$1 (d/190 pc)$^2$ & 1.05/3601\\ 
 &  &  & 2.88$_{-0.30}^{+0.35}$, CF=0.61$\pm$0.04 & &  &  &  \\
 
\hbox to 0.6in{\object{{4~Dra}(4050)}   \leaders\hbox to 0.2em{\hss.\hss}\hfill}&  \texttt{Tbabs$\times$pcfabs$\times$mkcflow} & $\cdots$ & 0.65$_{-0.05}^{+0.05}$ & 7.3$_{-0.8}^{+0.9}$ & 169$\pm$2 & 7$\pm$1 (d/190 pc)$^2$ & 1.01/1794 \\%Mdot 5.51521E-11 (-1.00093e-11,1.23588e-11)
 &  &  & 3$\pm$1, CF=0.40$\pm$0.05 & &  &  &  \\

\hbox to 0.6in{\object{{4~Dra}(4060)}   \leaders\hbox to 0.2em{\hss.\hss}\hfill}&  \texttt{Tbabs$\times$pcfabs$\times$mkcflow} & $\cdots$ & 1.16$\pm$0.08 & 11.5$\pm$0.7 & 683$\pm$2 & 30$\pm$1 (d/190 pc)$^2$ & 1.05/3610 \\
 &  &  & 3.1$\pm$0.4, CF=0.60$\pm$0.03 & &  &  &  \\

\hline

\enddata 
\tablenotetext{a}{Absorption column density (in units of 10$^{22}$ atoms cm$^{-2}$) and covering fraction (CF) of the partial absorber model \texttt{pcfabs}.}
\tablenotetext{b}{Indicates the value of the maximum temperature in the case of the cooling flow model \texttt{mkcflow}, $kT_{max}$, photon index of model \texttt{power-law}}
\tablenotetext{c}{Unabsorbed X-ray flux, in units of 10$^{-13}$ erg s$^{-1}$ cm$^{-2}$ in the 0.3-10.0 keV energy range.}
\tablenotetext{d}{Unabsorbed X-ray luminosity, in units of 10$^{31}$ erg s$^{-1}$ in the 0.3-10.0 keV energy range, scaled by the distances listed in Section \ref{sec:results}.}
\end{deluxetable}

\subsubsection{EG And}
\label{sec:reseg}

Our {\em Suzaku} observation took place at orbital phase $\phi$=0.93
\citep[using the ephemeris from][]{kolb04}, with the WD moving behind
the red giant wind (or $\phi$=0.17 if we use the orbital period of 481
days and the ephemeris from \citet{vogel1991}, i.e. the WD coming out
of partial eclipse). After visual inspection of the {\em Suzaku} spectrum, we verified that no obvious features were present and grouped the channels to have a minimum of 20 counts per bin, which allowed us to use $\chi^{2}$ statistics to assess the quality of the fit. 
Several pieces of evidence suggest that the accreting compact
  object is a non-magnetic WD and thus that the X-ray spectrum could be modeled as due to optically thin thermal emission. 
To our knowledge, periodic modulations at the WD spin have not been
detected either in optical or X-ray wavelengths, suggesting that
synchrotron emission from a strong magnetic field is not present. In
addition, \citep{kolb04} successfully model the UV emission with NLTE
atmospheric models for a low-mass WD.
We thus consider the best-fit model the one that consists of an absorbed, optically thin thermal plasma. The parameters are listed in Table \ref{tab:models}.

\subsubsection{4~Dra}% (=CQ~Dra)}
\label{sec:res4dra}
 
We analyzed the two {\em Suzaku} observations, ObsID 405035010 and 406041010 (hereafter 4050 and 4060, respectively; see Table \ref{table:info1}). Although in terms of $\chi^{2}_{\nu}$, an absorbed, non-thermal plasma plus a gaussian emission line model
(\texttt{Tbabs$\times$(power+gauss)}) 
fits the observed spectrum, the line centroid is at $\sim$6.67 keV with an equivalent width 0.17 keV, consistent with \ion{Fe}{25} transitions from a thermal plasma. We thus prefer a thermal origin for the observed X-ray emission.

We obtained acceptable fits with spectral models of optically thin thermal plasma emission in two different scenarios. In the first scenario, we simultaneously fit both observations linking the temperature of the optically thin thermal emission for both observations while allowing the absorption column to vary independently. This is a valid assumption if the X-ray emitting plasma arises in the post-shock region of the accretion disk boundary layer, and its temperature is set by the WD mass, which does not change between observations. In the second scenario, we modeled both observations independently. 

\begin{figure*}[ht!]
\begin{center}
  \includegraphics[scale=1.0]{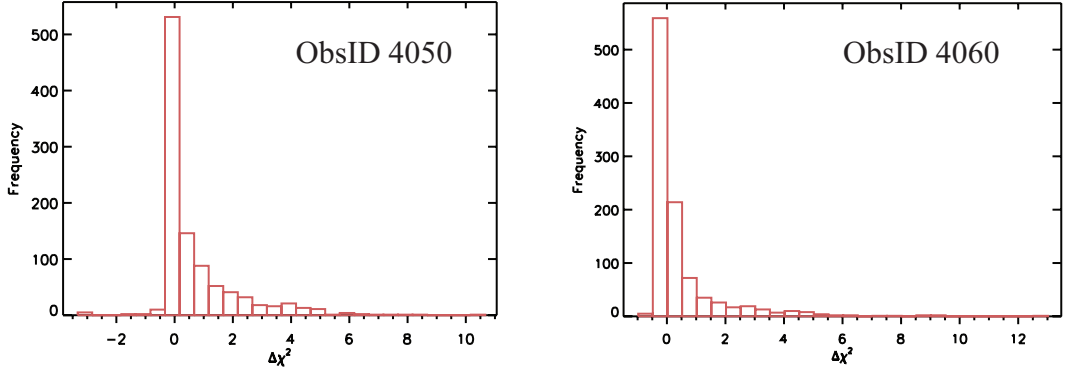}
 \caption{Histograms of the distribution of $\Delta\chi^{2}$ for ObsID 4050 and 4060 resulting from fitting 1,000 simulated spectra with
an optically thin thermal plasma observed through a simple absorber (model A) and observed through simple and partially covering absorbers (model B). Only 11\% of 1,000 randomly generated spectra produce a $\Delta\chi^{2}$ (i.e., difference in $\chi^{2}$ between the fit of model A and B) lower than zero and thus compatible with model A.} 
\label{fig:histo}
\end{center}
\end{figure*}

The first spectral model consists of an absorbed, single temperature plasma with a reduced Fe abundance (vapec\footnote{https://heasarc.gsfc.nasa.gov/xanadu/xspec/manual/XSmodelApec.html}). 
This model is formally acceptable for the two observations (see Table \ref{tab:models}).  However, in the case of 4060 there are significant residuals at energies below $\sim$1 keV, suggesting that a simple absorption model is not completely appropriate. The fact that some flux is detected at energies below $\sim$ 1 keV might be the footprint
of an absorber that partially covers the X-ray source. We thus added such an absorber to our spectral model, significantly improving the fit for both observations. To quantify the fit improvement by the addition of a partial absorber, we performed 1,000 simulations of both models following the LRT test shows that for 89\% of the simulations, the fit improves with the extra, partial covering absorber (see Fig. \ref{fig:histo}).

As an alternative, the second model consists of a multi-temperature cooling flow, again observed through both simple and partial covering absorbers.  
The X-ray spectrum was harder and the flux was higher for 4060 than for 4050. This is reflected in the higher N$_{H,22}$ (absorbing column in units of 10$^{22}$ cm$^{-2}$) and covering fraction obtained using the first scenario, and the higher temperature obtained for the second scenario.

\subsection{Timing analysis}
\label{sec:time}

Significant stochastic variability of the the X-ray flux on short (minutes to hours) time scales is a hallmark of accretion disks in binary systems. 
Periodic modulation, on the other hand, indicates that the accretion is channeled by an strong magnetic field \citep[e.g. Z~And][]{sokoloski06}. 
We searched for periodic variations in the X-ray flux from our 5 targets, using light curves binned at 16 s and the Lomb-Scargle algorithm. No periods with a 
p-value$\lesssim$0.3 were found for the sources in our sample.  We used the ratio between the observed and expected rms variabilities ($s$ and $s_{exp}$, respectively) to
quantify stochastic variations in the light curves. We binned the light curves at integer multiples of the XIS readout time: 16, 160, 1600, and 3680 seconds, and calculated the ratio $s/s_{exp}$ (see Table \ref{table:timing}).  Only 4~Dra showed significant variability, with amplitudes as high as 40\%; the strongest flickering was present on time scales of hours, but there was also intense flickering on time scales of minutes, consistent with previous determinations \citep{wheatley03}. 

\renewcommand{\arraystretch}{1.5}
\begin{deluxetable}{ccccccc}
\tabletypesize{\footnotesize}
\tablecolumns{7}
\tablewidth{0pt}
\tablecaption{Timing analysis\tablenotemark{a} \label{tab:time} \label{table:timing}}
\tabletypesize{\scriptsize}
\tablehead{
\colhead{Object} & \colhead{Bin size [s]} & \colhead{s$_{exp}$} & \colhead{s} & \colhead{Ratio} & 
%\colhead{s$_{exp}$/average[\%]} & \colhead{s/average[\%]}
}
\startdata
\hbox to 0.9in{Hen~3-1591\leaders\hbox to 0.8em{\hss.\hss}\hfill} & 160 & 0.05 & 0.05 & 1.02 \\ %& 52 & 53  \\
\hbox to 0.9in{4~Dra(4050)\leaders\hbox to 0.8em{\hss.\hss}\hfill} & 160 & 0.05 & 0.17 & 3.32 \\ %& 15 & 50 \\
\hbox to 0.9in{4~Dra(4060)\leaders\hbox to 0.8em{\hss.\hss}\hfill} & 160 & 0.08 & 0.39 & 4.72 \\ %& 8 & 38 \\
 \hline
\hbox to 0.9in{CD-28~3719\leaders\hbox to 0.8em{\hss.\hss}\hfill}& 1600 & 0.01 & 0.01 & 1.08 \\ %& 47 & 50 \\
\hbox to 0.9in{Hen~3-1591\leaders\hbox to 0.8em{\hss.\hss}\hfill} & 1600 & 0.02 & 0.03 & 1.30 \\ %& 20 & 26  \\
\hbox to 0.9in{4~Dra(4050)\leaders\hbox to 0.8em{\hss.\hss}\hfill} & 1600 & 0.02 & 0.14 & 7.27 \\ %& 6 & 43 \\
\hbox to 0.9in{4~Dra(4060)\leaders\hbox to 0.8em{\hss.\hss}\hfill}& 1600 & 0.03 & 0.33 & 10.26 \\ %& 3 & 32 \\
 \hline
\hbox to 0.9in{CD-28~3719\leaders\hbox to 0.8em{\hss.\hss}\hfill}  & 3680 & 0.01 & 0.01 & 1.24 \\ %& 39 & 48 \\
\hbox to 0.9in{EG~And\leaders\hbox to 0.8em{\hss.\hss}\hfill}& 3680 & 0.01 & 0.01 & 1.18 \\ %& 52 & 62  \\
\hbox to 0.9in{4~Dra(4050)\leaders\hbox to 0.8em{\hss.\hss}\hfill} & 3680 & 0.01 & 0.14 & 9.05 \\ %& 4 & 40 \\
\hbox to 0.9in{4~Dra(4060)\leaders\hbox to 0.8em{\hss.\hss}\hfill} & 3680 & 0.02 & 0.32 & 13.71 \\ %& 2 & 30 \\

 \enddata
\tablenotetext{a}{The values $s$ and $s_{exp}$ have units of c/s.  
%The quantity ``average'' in the final columns refers to the average count rate in the same units. 
We list only those bin sizes for which $s \gtrsim s_{exp}$.} 
\end{deluxetable}

\section{Discussion}
\label{sec:disc}

The X-ray emission in all five symbiotic stars observed with {\it Suzaku} is consistent with a thermal origin, indicating the presence of shocked gas. We can use the derived temperature of this gas to infer the physical mechanism responsible for its production.  In the case of X-rays from colliding winds ($\beta$-type emission), the post-shock temperature is related to the velocity difference between the two winds fast and slow ($(4/3)\times(v_{fast} - v_{slow})= v_{shock} $) by the Rankine-Hugoniot strong shock conditions: 

\begin{equation}
%v_shock = (4/3)\times(v_{fast} - v_{slow})
%v_{fast} - v_{slow} 
v_{shock} = (4/3)\times \sqrt{\frac{16 k T_{s}} {3\mu m_{H}}}
\end{equation}

\noindent where $T_{s}$ is the post-shock temperature, $\mu$ the mean
molecular weight, and $m_{H}$ is the mass of a hydrogen atom.   

Temperatures higher than kT $\sim$ 1--2 keV are difficult to explain
with this mechanism, since there is little evidence of outflows with
velocities $>$ 1000 km s$^{-1}$ in symbiotic stars.   If the emission
instead originates in the accretion disk boundary layer (the assumed
origin of $\delta$-type emission), then the temperature of the shocked
gas is set by the gravitational potential of the WD, and can
be used to infer the mass of this.  Assuming accretion
through a disk, then the velocity at the boundary layer is set by the
inner Keplerian condition, and a lower limit on the mass can be calculated as follows:

\begin{equation}
M_{WD} \ge \frac{16 T_{s} k R_{WD}}{3 G \mu m_{H}},
\end{equation}

\noindent where $G$ is the gravitational constant and $R_{WD}$ is the radius of the WD, calculated using the mass-radius relation of \citet{pringle}.
The mass derived in this way should be considered only a lower limit because additional cooling mechanisms could be at play in the boundary-layer region.  One source of cooling that could be relevant for these systems is Compton cooling via UV photons from any optically thick portion of the boundary layer. This mechanism has been invoked in dwarf novae in outburst \citep{fertig} and in the recurrent nova RS Oph after its 2006 outburst \citep{nelson}.

\subsection{CD-28~3719} 
%\label{sec:disccd-28}

CD-28~3719 was detected in the X-rays for the first time in a short pointing observation with {\em Swift}, and in Paper I we classified its spectrum as $\delta$-type. The {\em Suzaku} spectrum also shows the presence of hard, optically thin thermal X-ray emission observed through strong absorption, still compatible with a $\delta$-type symbiotic. Using the plasma temperature from the cooling flow model to constrain the WD mass, we obtain $M_{WD} > 0.6 M_{\odot}$. CD-28~3719 seems to be the most stable source in our sample.  

\subsection{Hen~3-1591}
%\label{sec:dischen31591}
Only two objects were classified by \citet{murset97} as $\gamma$-type symbiotics based on their emission detected with {\em ROSAT}: GX~1+4 and Hen~3-1591. These observations had a limited signal-to-noise ratio, but \citet{murset97} speculated that these sources harbored neutron stars as accretors. Lately, this type of symbiotic was named ``symbiotic X-ray binaries" by \citet{masetti07}.

Our best-fit model for \object{Hen~3-1591}, which consists of a thermal plasma rather than a power-law, points to the presence of a WD as the accreting compact object instead of a neutron star, i.e. \object{Hen~3-1591} is a WD symbiotic instead of a symbiotic X-ray binary as previously thought. Evidence supporting this scenario also comes from the optical spectrum, which shows many emission lines of a variety of ions, similar to that of a planetary nebula, where the strong UV radiation field from the WD photoionizes the surrounding nebula. \citet{hedrick04} detected flickering behavior in the B band emission from this object, which suggests that the blue light is from an accretion disk.  Hen~3-1591 belongs to a rare subclass of $d^{\prime}$-type yellow symbiotics, with a dusty IR continuum and located in the Galactic disk, so low metallicity is not implicated.  They are interpreted as systems in which the hot component has recently evolved from the AGB to the WD stage. In this interpretation, the dust 
is from the mass lost by the AGB star, and the nebulosity is in fact the planetary nebula (PN), and neither is due to the present-day giant (see Jorissen et al. 2005 and references therein). This symbiotic system could therefore be in some ways comparable to the well-known classical novae GK~Per \citep{bode87} and V458~Vul \citep{wesson08}, both of which have been  claimed to be classical novae inside  PNe. These systems are all quite special if the PNe phase is as short as thought \cite[10-20 kyr;][]{badenes15}.

Hen~3-1591 also shows the {\em barium syndrome}, i.e. overabundances of $s$-process elements and the presence of singly-ionized barium, which cannot be explained unless the red giant is part of a binary system \citep{jorissen92, jorissen05}. The now-observed red giant had its photosphere polluted by $s$-processed elements by the previously-AGB companion, which now should be a WD. Thus, our high signal-to-noise, broadband {\em Suzaku} spectrum adds more support to the presence of a WD or hot subdwarf in \object{Hen~3-1591}. If we use the maximum plasma temperature from the cooling flow model to constrain the WD mass, we obtain M$_{WD} \ge $0.45 M$_{\odot}$.

The high temperature of the plasma strongly suggests that in Hen~3-1591 the X-ray emission arises in an accretion disk boundary layer instead of a colliding wind region. The strong-shock condition implies wind speeds of around 3,000 km s$^{-1}$ for the observed temperatures, and such high-speed outflows or winds have not been detected in Hen~3-1591 or almost any other symbiotic. The lack of UV data does not allow us to use the ratio of UV and X-ray fluxes as a proxy for the optical depth of the boundary layer. The decrease in temperature and luminosity between the {\em ASCA} and {\em Suzaku} observations, however, suggests that the optical depth of the X-ray emitting plasma changed, being higher during the {\em Suzaku} observation.  {\em Suzaku} might thus have observed a smaller optically thin portion of the boundary layer.

\subsection{Hen~3-461}
%\label{sec:dischen3461}

Hen 3-461 was discovered in X-rays with {\em Swift} during a short pointing observation (Paper I). The high temperature and absorption obtained from modeling those data, the hardness ratio (defined as the ratio of count rates in the 2.4-10/0.3-10 keV ranges) and the presence of significant flickering in the UV, all suggested that the X-ray emission originated in the boundary layer of the accretion disk and led us to classify it as a $\delta$-type source. 

Our {\em Suzaku} observation indicates two important changes since the {\em Swift} observation: increased absorption towards the hard X-rays from the boundary layer, and the appearance of a new, softer $\beta$-type component below 2 keV. Thus, the source show us a $\beta/\delta$ type spectra.

The intrinsic X-ray luminosity decreased by about 30\% between the {\em Swift} observation in 2010 and the {\em Suzaku} observation in 2012 December.  Although no contemporaneous UV data are available, GALEX (NUV) observations taken one year before our {\em Suzaku} observation indicate that F$_{UV}$=1.86$\times$10$^{-12}$ erg cm$^{-2}$ s$^{-1}$, so that F$_{UV}$/F$_{X} \gtrsim$ 0.7 (we quote a lower limit as the reddening for this source is unknown).  Thus the accretion disk boundary layer seems to be still in the optically thin regime.  In Paper I we proposed a scenario in which the soft emission of $\beta/\delta$-type objects could be related to a colliding wind region or jets. We could be witnessing the injection of new flows into a colliding wind region. The equivalent width of the (unresolved) Fe K$\alpha$ region of around 400 eV resembles 
the values found by \citet{mukai07} on the well-known jet-source with a two component X-ray spectra, CH~Cyg.

\subsection{EG~And}
%\label{sec:disceg}

In the V band, EG~And is one of the brightest symbiotic systems. Its distance is 512$\pm$168 pc \citep{vanleuween07}. The periodic photometric modulation indicates an orbital period of 470 days with an inclination of 82$^{+8}_{-4.5}$ degrees \citep{kolb04}, making EG~And an eclipsing symbiotic binary. 

Based on the high temperature ($kT = 3$~keV) found for the X-ray emitting plasma (see Table \ref{tab:models}), we can estimated a M$_{WD}\approx$0.4 M$_{\odot}$ and rule out colliding winds as the source of X-rays in EG~And. This source was identified as a $\beta$-type symbiotic by \citet{murset97}. Those authors derived a temperature for the X-ray emitting plasma of $kT =1.3\pm$0.5 keV using {\em ROSAT} PSPC data ---  much lower than the temperature from our model of the {\em Suzaku} spectrum. The difference is most likely due to the lack of sensitivity above 2 keV of the {\em ROSAT} spectrum.  The velocity implied by the best fit to the {\em Suzaku} spectrum is $\sim$1200km s$^{-1}$ which is significantly faster than the highest velocity line features ($\sim$700 km s$^{-1}$, C IV 1548, 1550\AA~\rm  absorption features) found in UV spectra from FUSE and STIS by \cite{crowley08}, and difficult to explain with outflows from a low mass WD like the one in EG~And by \citet[M$_{WD}\approx$0.4 M$_{\
odot}$;][]{kolb04}.

If the hot-component luminosities reported in the literature \cite[16-400 $L_{\odot}$;][]{vogel92,kolb04} are due to an accretion-powered WD, then the implied accretion rates are in the range 10$^{-8}$ to 10$^{-7}$ M$_{\odot}$ yr$^{-1}$ for a 0.4 M$_{\odot}$ WD.  This is squarely in the regime where the boundary layer is expected to be optically thick according to models by \cite{popham95}. The X-ray luminosity is orders of magnitude lower than $L_{hot}$, suggesting that the X-rays are produced in a region that remains optically thin at the outer surface of the mostly optically thick boundary layer. Although we expect X-ray emission due to accretion to be highly variable, EG~And is very faint, and we do not detect enough photons to be sensitive to low-level variability ($s_{exp}/<\rm count rate \rm> \sim$ 50\%).  

\subsection{4~Dra}
%\label{sec:dischen4Dra}

Although 4 Dra was classified as a triple system consisting of 4~Dra(A) + CQ~Dra(Bab) by \citet{reimers85}, X-ray data obtained with {\em ROSAT} led \citet{wheatley03} to suggest that the X-ray emission is consistent with the presence of a WD accreting from the wind of a red giant. IUE data led \citet{skopal} to the same conclusion after
fitting the SED. More evidence against the triple-system nature of 4~Dra comes from the analysis of the broad wings superimposed upon narrow emission lines in FUSE spectra by \citet{froning12}, who found the FUSE spectra to be similar to other FUSE spectra of confirmed symbiotic stars. 
Based on the previous studies and current results, when modeling the {\em Suzaku} spectra, we considered this source to be a symbiotic system with an accreting WD. 

This source is one of the two systems with Hipparcos distances in our sample, having $d$=190$\pm$17 pc \citep{vanleuween07}. Its optical brightness presents irregular variations of about 0.1 mag in V \citep{eggen67}, and radio flux variations also occur on time scales of weeks to months and maybe shorter than hours to days \citep{brown87}. 

Our {\em Suzaku} observations of 4~Dra shows strong variability, a high $kT$, and X-ray luminosities in the range of 0.01 L$_{\odot}$. If considered along with previous UV luminosity estimates of L$_{UV} \sim$10 by \cite{skopal}, our findings suggest that 4~Dra is an accretion-powered symbiotic (i.e., no quasi-steady shell burning on the surface of the WD) in which mass is transferred at $\sim$10$^{-8}$ M$_{\odot}$ yr$^{-1}$. 

In such a case, the bulk of the boundary layer luminosity is emitted in the soft X-rays or EUV and is unobservable at Earth, while the remaining optically thin part of the boundary layer emits faint hard X-rays \citep{patterson85}. 
Our inference that the boundary layer is optically thick enables us to estimate the mass of the WD. \cite{popham95} suggest that the accretion rates needed to produce a predominantly optically thick boundary layer around WDs of masses 1.0, 0.8, and 0.6 M$_{\odot}$ are greater than 10$^{-7}$, a few times 10$^{-8}$, and $\sim$10$^{-8}$ M$_{\odot}$ yr$^{-1}$, respectively.  Taking the UV flux at the time of the {\em Suzaku} observation to be comparable to the $\sim$10 L$_{\odot}$ detected by IUE, the BL being optically thick suggests that the WD mass is less than about 0.6 M$_{\odot}$ (the IUE luminosity corresponds to an accretion rate of $\sim$10$^{-8}$ M$_{\odot}$ yr$^{-1}$). 

We emphasize that our conclusions about the optical depth of the accretion disk boundary layer and its implications for the WD mass depend on the assumed viscosity parameter, $\alpha$, and other assumptions used in the models described by \citet{popham95}. As noted in Paper I, a 30\% change in $\alpha$ leads to a factor of a few change in the threshold accretion rate above which the boundary layer is expected to become optically thick. Allowing for this, the WD in 4~Dra probably has a mass less than about 0.7 M$_{\odot}$, still a low-mass WD.

Both fitting approaches discussed in Section \ref{sec:res4dra} yield equally possible scenarios for 4~Dra. In the case that both spectra are modeled independently, the measured temperatures could be different because the optical depth of the emitting region changed, i.e. during the earliest observation the boundary layer was more optically thick, and the measured temperature and fluxes of the X-ray emitting plasma were lower. On the other hand, if we assume that during both observations the plasma temperature was the same (the optical depth of the boundary layer did not change between observations), we still obtain acceptable fits with an increase in the amount of absorbing material and intrinsic X-ray luminosity between the two observations, perhaps due to an increase in the accretion rate.  

From the values of absorbing column listed in Table~\ref{tab:models}, we conclude that the absorbing material is located relatively near the WD. 
The orbital solution for 4~Dra was studied by \citet{famaey09}, who used the radial velocities of the M giant to find a 1703$\pm$3 day period, an eccentric orbit ($e$=0.3), and T$_{0}$[MJD]= 53204$\pm$19. The {\em ROSAT} observations thus occurred at orbital phases $\phi$=0.33, 0.15 and 0.61, while {\em Suzaku} observed at $\phi^{4050}$=0.23 and $\phi^{4060}$=0.57. If the X-ray absorbing material is tied to the orbital motion of the WD, the absorbing columns obtained from the fits of the {\em Suzaku} observations should not be very different, given that during ObsID 4050 the system was near quadrature, while during ObsID 4060 the WD was in inferior conjunction. 
In that picture, the second observation (4060) taken during the inferior conjunction should even be somewhat less absorbed, which is not the case. It is therefore probably intimately connected with the physics of accretion rather than to the orbit and the red giant companion.

\section{Conclusions}
\label{sec:concl}

We analyzed the deep, broad-band {\em Suzaku} observations of five symbiotic stars previously known to be X-ray sources. In contrast to observations obtained with {\em ROSAT}, which had limited  energy coverage, or those obtained with {\em Swift}, which had limited continuous observing time, the spectra obtained with {\em Suzaku} allowed us to better unveil the origin of the X-ray emission in symbiotics. The high temperatures of the plasma of a few keV (see Table \ref{tab:models}) imply shock speeds of more than a thousand km~s$^{-1}$ (assuming strong shock conditions). Given that such high-speed outflows are not typically observed in symbiotic stars, this finding strongly suggest that the emission originates in an accretion disk boundary layer rather than a colliding-wind region. 

The high temperatures are roughly consistent with shock-heated gas at the virial temperature in the deepest portion of the WD potential well, suggesting that Compton cooling of the plasma in the boundary layer is not an important source of cooling.

As discussed in Paper I, Compton cooling of the post-shock region could be important if there is a source of abundant UV photons, high local accretion rate, and a relatively massive WD. The low UV luminosities from the sources presented here indicate that shell-burning on the WD surface is not important, and thus the conditions for prevalent Compton cooling are not satisfied.  

In Paper I, we proposed a new classification system for the X-ray emission from symbiotic stars, in which the category $\delta$ was assigned to those sources with a high absorbing column  ($\sim$ a few 10$^{22}$ cm$^{-2}$) and  thermal X-ray emission with energies above 2.4 keV. The fraction of X-ray emission radiated in this hard regime (see Fig. 4 in Paper I) may vary between different observations of the same source.

We calculated the hardness ratios (HR) as defined in Paper 1 using the {\em Suzaku} data or folding our best-fit models through the {\em Swift/XRT} responses. In both cases, we found that except for CD~-28~3719, all sources observed with {\em Suzaku} have HR $\lesssim$1.  In fact, for Hen~3-461 and CD~-28~3719, the HRs changed dramatically between the first observations with {\em Swift} and the observations with {\em Suzaku}. Figure 4 in Paper 1 showed that during the {\em Swift} observations, all $\delta$-type sources had HR$\gtrsim$ 1. The fact that we observed most sources to have HR$\lesssim$ 1 indicates that while the $\delta$ components in the \citeauthor{Paper1} sample were all heavily absorbed, this is not universally true of all $\delta$ components; in this study, Hen~3-1591, EG~And, and 4~Dra were all lightly absorbed and detected below 1 keV. 

When compared with earlier data, all sources show changes in their intrinsic X-ray flux and N$_{H}$. Taking $\delta$-type X-ray emission to originate in the accretion disk boundary layer, long term changes in the X-ray flux are mostly related to changes in the accretion rate onto the WD, while changes in the soft X-rays can also be caused by variations in the amount of absorbing material. 

The days-to-week time scale changes in N$_H$ observed in the $\delta$-type prototype RT~Cru \citep{luna07, luna10} suggest that the absorber in that system is located close to the WD. However, it is unclear if these changes are related to the amount of material flowing through the accretion disk, and if so, how. In this study, we witness that although high flux states have high N$_{H}$ in the cases of CD~-28-3719 and 4~Dra,  in the case of Hen~3-461, N$_{H}$ is higher now while F$_{X}$ is lower than when it was observed with {\em Swift} (see Table 2 in Paper I).

Of the five sources observed with {\em Suzaku}, CD~-28~3719 retains its previous classification as a $\delta$-type source (see Paper I), as derived from {\em Swift} observations. The proposed WD nature of the compact object in Hen~3-1591 and the temperature of the X-ray emitting plasma suggest that it should now be considered a $\delta$-type source instead of $\gamma$-type source \citep[those symbiotics with neutron stars as compact objects; see][]{murset97}. The likely presence of a new soft component in the X-ray spectrum of Hen~3-461 encourages us to propose a $\beta/\delta$-type classification for that object. No X-ray spectral type has previously been proposed for 4~Dra, and given the results obtained from our spectral fit, we advocate for a $\delta$-type categorization. Finally, the low X-ray temperature derived for EG And which is consistent with the low mass estimate based on spectroscopy by \citet{kolb04} within the uncertainties on kT and the broadband energy coverage now shows that EG~And 
should be considered a $\delta$-type instead of $\beta$-type source as originally proposed by \citet{murset97}.

\acknowledgments
NEN acknowledges Consejo Nacional de Investigaciones Cient\'ificas y T\'ecnicas, Argentina (CONICET) for the Postdoctoral Fellowship. GJML and NEN acknowledge funding from PIP D-4598/2012, ANPCYT-PICT 0478/14 and Cooperaci\'on Internacional \#D2771 from Consejo Nacional de Investigaciones Cient\'ificas y T\'ecnicas, Argentina. KM acknowledges support by NASA through ADAP grant NNX13AJ13G. JLS acknowledges support by NASA through ADAP grant NNX15AF19G. This research has made use of data obtained from the {\em Suzaku} satellite, a collaborative mission between the space agencies of Japan (JAXA) and the USA (NASA) and the VizieR catalogue access tool, CDS, Strasbourg, France. The original description of the VizieR service was published in \citet{vizier00}.

{\it Facilities:} \facility{{\em Suzaku}}.


\begin{thebibliography}{}
\bibitem[Abdo et al.(2010)]{abdo2010} Abdo, A. A., Ackermann, M., Ajello, M., Atwood, W. B. et al., 2010, Science, 329, 817
\bibitem[Ackermann et al.(2014)]{v407cyg} Ackermann, M., Ajello, M., Albert, A.,et al., 2014, Science, 345, 554
\bibitem[Badenes et al.(2015)]{badenes15} Badenes, C., Maoz, D., Ciardullo, R. 2015, 2015arXiv150201015B
\bibitem[Belczy{\'n}ski et al.(2000)]{bel00} Belczy{\'n}ski, K., Miko{\l}ajewska, J., Munari, U., Ivison, R. J., Friedjung, M. 2000, \aaps, 146, 407
\bibitem[Bode et al.(1987)]{bode87} Bode, M. F., Roberts, J. A., Whittet, D. C. B., Seaquist, E. R., Frail, D. A., 1987, Nature, 329, 519
\bibitem[Brocksopp et al.(2004)]{brocksopp04} Brocksopp C., Sokoloski, J. L., Kaiser, C., et al. 2004, \mnras, 347, 430
\bibitem[Brown(1987)]{brown87} Brown, A. 1987, \apjl 312, L51
\bibitem[Cash(1979)]{cash} Cash, W. 1979, \apj, 228, 939
\bibitem[Crocker et al.(2001)]{crocker01} Crocker, M. M., Davis, R. J., Eyres, S. P. S., et al. 2001, \mnras, 326, 781
\bibitem[Crowley et al.(2008)]{crowley08} Crowley, C., Espey, B. R., McCandliss, S. R. 2008, \apj, 675, 711
\bibitem[Eggen (1967)]{eggen67} Eggen, O. J. 1967, Astrophysical Supplement Series, 14, 307
\bibitem[Famaey et al.(2009)]{famaey09} Famaey, B., 2009, A\&A, 498, 627
\bibitem[Fertig et al.(2011)]{fertig} Fertig, D., Mukai, K., Nelson, T. \& Cannizzo, J. K. 2011, \pasp, 123, 1054
\bibitem[Froning et al.(2012)]{froning12} Froning, C. S. et al., 2012, ApJS, 199, 7 
%\bibitem[Hansen \& Kawaler(1994)]{hansen94} Hansen, Carl J., Kawaler, Steven D. 1994, Stellar Interiors: Physical Principles, Structure, and Evolution (1th Ed.; New York, NY, Springer)
\bibitem[Hedrick \& Sokoloski(2004)]{hedrick04} Hedrick, C., Sokoloski, J. 2004, BAAS, 36, 1525
\bibitem[Kellogg et al.(2007)]{kellogg07} Kellogg, E., Anderson, C., Korreck, K., et al., \apjl, 664, 1079
\bibitem[Jorissen \& Mayor(1992)]{jorissen92} Jorissen, A., Mayor, M., International Astronomical Union Supplement, 151, 407
\bibitem[Jorissen et al.(2005)]{jorissen05} Jorissen, A., Zacs, L., Udry, S., Lindgren, H. \& Musaev, F.A. 2005, \aap, 441, 1135
\bibitem[Kennea et al.(2009)]{kennea09} Kennea, J.,Mukai, K., Sokoloski, et al., 2009, \apj, 701, 1992
\bibitem[Kenyon(1986)]{kenyon09} Kenyon, S. J. 2009, The Symbiotic Stars, 1st edition, Cambridge University Press
\bibitem[Kolb et al.(2004)]{kolb04} Kolb, K. M., J. Sion, E. M., Miko{\l}ajewska, J. 2004, \aj, 128, 1790
\bibitem[Koyama(2007)]{koyama07} Koyama, K., Tsunemi, H., Dotani, T., et al. 2007, \pasj, 59, 23
\bibitem[Luna \& Sokoloski(2007)]{luna07} Luna, G. J. M. \& Sokoloski, J. L. 2007, \apj, 671, 741
\bibitem[Luna et al.(2008)]{luna08} Luna, G. J. M., Sokoloski, J., Mukai, K., 2008, \ Astronomical Society of the Pacific, 401, 342
\bibitem[Luna et al.(2010)]{luna10} Luna, G.J.M., Sokoloski, J., Mukai, K. \& Nelson, T. 2010, Astronomer's Telegram \#3053, http://www.astronomerstelegram.org/?read=3053
\bibitem[Luna et al.(2013)]{Paper1} Luna, G. J. M., Sokoloski, J. L., Mukai, K. \& Nelson, T., 2013, \aap, 559, 6
\bibitem[Masetti et al.(2007)]{masetti07} Masetti, N.; Rigon, E.; Maiorano, E. et al.,  2007a, \aap, 464, 277 %4U 1954+31
\bibitem[M\"urset et al.(1997)]{murset97} M\"urset, U., Wolf, B. \& Jordan, S. 1997, \aap, 319, 201
\bibitem[Mukai et al.(2007)]{mukai07} Mukai, K., Ishida, M.; Kilbourne, C., et al., 2007, PASJ, 59, 177
\bibitem[Nelson et al. (2011)]{nelson} Nelson, T., Mukai, K., Orio, M., Luna, G. J. M. \& Sokoloski, J. L, 2011, \apj, 737, 7
\bibitem[Nu\~nez at al.(2014)]{nunez14} Nu\~nez, N. E., Luna, G. J. M., Pillitteri, I., Mukai, K., 2014, \aap, 565, A82
\bibitem[Ochsenbein, Bauer \& Marcout(2000)]{vizier00} Ochsenbein, F.; Bauer, P. \& Marcout, J. 2000, A\&ASS,143, 23
\bibitem[Patterson \& Raymond(1985)]{patterson85} Patterson, J.; Raymond, J. C. 1985, \apj, 292, 550
\bibitem[Popham \& Narayan(1995)]{popham95} Pophan, R. \& Narayan, R. 1995, \apj, 442, 337
\bibitem[Pringle \& Webbink (1975)]{pringle} Pringle, J. E. \& Webbink, R.F. 1975, \mnras, 172, 493
\bibitem[Protassov et al.(2002)]{protassov} Protassov, R., van Dyk, D. A., Connors, A. et al., 2002, \apj,571,545
\bibitem[Reimers(1985)]{reimers85} Reimers, D. 1985, \aap, 142L, 16
\bibitem[Seaquist \& Taylor(1990)]{seaquist90} Seaquist, E. R., Taylor, A. R, 1990, \apj, 349, 313 
\bibitem[Seaquist et al.(1993)]{seaquist93} Seaquist E. R., Krogulec, M., Taylor, A. R, 1993, \apj, 410, 260
\bibitem[Skopal(2005)]{skopal} Skopal A., 2005, A\&A, 440, 995
\bibitem[Sokoloski et al.(2006)]{sokoloski06} Sokoloski, J. L., Kenyon, S. J., Espey, B. R., et al., \apj, 636, 1002
\bibitem[Tanaka et al.(1994)]{tanaka94} Tanaka, Y., Inoue, H. \& Holt, S.S. 1994, \pasj, 46, L37
\bibitem[van Leuween (2007)]{vanleuween07} van Leuween, F. 2007, \aap, 474, 653
\bibitem[Verner \& Yakovlev(1996)]{verner} Verner, D. A. \& Yakovlev, D. G. 1995, \aaps, 109, 125
\bibitem[Vogel (1991)]{vogel1991} Vogel, M., 1991, A\&A, 249, 173
\bibitem[Vogel et al.(1992)]{vogel92} Vogel, M., Nussbaumer, H., Monier, R. 1992, \aap, 260, 156
\bibitem[Wesson et al.(2008)]{wesson08} Wesson, R., Barlow, M. J., Corradi, et al. 2008, \apjl, 688, L21
\bibitem[Wheatley et al.(2003)]{wheatley03} Wheatley, P. J., Mukai, K., de Martino, D. 2003, \mnras, 346, 855
\bibitem[Wilms et al.(2000)]{wilms} Wilms, J., Allen, A. \& McCray, R. 2000, \apj, 542, 914

\end{thebibliography}
\end{document}